\documentclass{article}
\usepackage{spconf,amsmath,graphicx}

\usepackage{multirow}
\usepackage{array}
\usepackage{hhline}
\usepackage{graphicx}
\usepackage{graphics}
\usepackage{comment}
\usepackage{adjustbox}
\usepackage{tabularx,booktabs}
\usepackage[utf8]{inputenc}
\usepackage{mhchem,multirow,calc,array,amsmath,booktabs}
\usepackage{hyperref}
\usepackage{booktabs}
\newlength\mylenA
\newlength\mylenB
\newlength\mylenC
\newlength\mylenD

\title{MiniSUPERB: Lightweight Benchmark for Self-supervised Speech Models}
\name{Yu-Hsiang Wang, Huang-Yu Chen, Kai-Wei Chang, Winston Hsu, Hung-yi Lee}
\address{College of Electrical Engineering and Computer Science, National Taiwan University, Taiwan}
\copyrightnotice{979-8-3503-0689-7/23/\$31.00~\copyright2023 IEEE}
\begin{document}
%\ninept
%
\maketitle
\begin{abstract}
SUPERB was proposed to evaluate the generalizability of self-supervised learning (SSL) speech models across various tasks. However, it incurs high computational costs due to the large datasets and diverse tasks. In this paper, we introduce MiniSUPERB, a lightweight benchmark that efficiently evaluates SSL speech models with comparable results to SUPERB but lower computational costs significantly. We carefully select representative tasks, sample datasets, and extract model representations offline. Our approach achieves a Spearman's rank correlation of 0.954 and 0.982 with SUPERB Paper and SUPERB Challenge, respectively. Additionally, we reduce the computational cost by 97\% in terms of Multiply-ACcumulate operations (MACs). Furthermore, we evaluate SSL speech models in few-shot scenarios and observe significant variations in their performance. To our knowledge, this is the first study to examine both the computational cost of the model itself and the cost of evaluating it on a benchmark. \footnote{Our code is available at \url{https://github.com/Comet0322/MiniSUPERB}}

\end{abstract}

\section{Introduction}
\label{sec:intro}

Self-supervised learning (SSL) has gained increasing attention in recent years as a promising approach for learning informative representations without human-annotated labels. One of the main advantages of SSL is that it can effectively leverage large amounts of unlabelled data, which is often easier to obtain than labeled data. SSL has shown impressive performance across various domains, including computer vision \cite{DBLP:conf/cvpr/00050XWYF22,DBLP:conf/iclr/Bao0PW22}, natural language processing \cite{DBLP:conf/naacl/PetersNIGCLZ18,DBLP:conf/naacl/DevlinCLT19}. In speech processing, SSL has been applied to a wide variety of tasks, including but not limited to speaker recognition \cite{DBLP:conf/icassp/LiuYCHL20}, speech recognition \cite{DBLP:conf/interspeech/SchneiderBCA19,DBLP:conf/icassp/LingLSK20}, and emotion recognition \cite{DBLP:conf/interspeech/PascualRSBB19}, achieving state-of-the-art performances.

To evaluate the efficacy of representation learned from SSL speech models, SUPERB (Speech processing Universal PERformance Benchmark)~ \cite{yang21c_interspeech} was proposed. SUPERB comprises dozens of speech processing tasks that test the universality of SSL speech models. It has proved to be instrumental in evaluating SSL models and promoting research in this field. However, each task in SUPERB requires a considerable amount of computation and time to complete. Consequently, evaluating a single SSL model on SUPERB can be a time-consuming process. Moreover, developing novel SSL speech models that require extensive evaluations during the development phase can be even more challenging.

To address this issue, we present MiniSUPERB, a light-weight benchmark for SSL speech models that aims to expedite the evaluation of speech SSL models with significantly reduced computation and time costs. MiniSUPERB can be seen as a proxy for SUPERB. It provides a representative subset of tasks that can be employed to verify model performance during development, and researchers can subsequently use SUPERB for a comprehensive evaluation and ranking once their models are sufficiently developed.

To enhance the effectiveness of SUPERB in producing consistent evaluation outcomes while minimizing computational costs, we have implemented four significant improvements. Firstly, we have chosen a representative subset of tasks for evaluation. Secondly, we have sampled the dataset to reduce the computation and storage requirements. Thirdly, we have pre-extracted representations to expedite the training process. Lastly, we have simplified the downstream model to minimize computation and training time.

Our experiments show that the ranking obtained from MiniSUPERB highly correlates with that obtained from SUPERB, with a 0.982 Spearman's rank correlation coefficient (Spearman's $\rho$). Moreover, we reduce the computational cost by 97\% and significantly shorten the evaluation time. To the best of our knowledge, this is the first work that studies the computational cost of the benchmark itself and the model being evaluated at the same time. MiniSUPERB provides researchers with a quick and efficient way to evaluate their SSL models in speech processing, avoiding unnecessary computation and time costs.

\section{Related works}
\subsection{Self-supervised speech model}
Recently, there has been an increased focus on self-supervised learning (SSL) among researchers in the speech processing field. This involves the use of reconstruction, masked prediction, and contrastive learning techniques on large, unlabeled corpora. For instance, HuBERT \cite{DBLP:journals/taslp/HsuBTLSM21} utilizes iterative prediction of speech feature cluster centers, while Wav2vec 2.0 \cite{DBLP:conf/nips/BaevskiZMA20} and CPC \cite{DBLP:journals/corr/abs-1807-03748} employ contrastive prediction learning. These SSL speech models can serve as feature extractor, which extracts informative representations that can benefit downstream tasks. The typical speech representation learning paradigm~\cite{mohamed2022self} consists of two stages: (1) pre-training an upstream model and (2) fine-tuning downstream models on downstream tasks. Various benchmarks have been suggested to assess the quality of the speech representations extracted by an upstream model, with SUPERB being the most commonly used benchmark.

\subsection{SUPERB benchmarks}
SUPERB is a benchmark that assesses the quality of representation encoded by SSL speech models through various speech processing tasks. Different versions of SUPERB have been developed to evaluate these models in different aspects, including semantic, speaker, paralinguistic, and generation aspects. The most common protocol in SUPERB involves fixing the parameters of a pre-trained upstream model, which acts as a feature extractor that collaborates with task-specific downstream models to perform various speech processing tasks. This protocol has proven useful in evaluating the quality of representation generated by the upstream model. In the following, we briefly introduce the previous three versions of SUPERB. Readers are encouraged to read the original papers for more detail.

\textbf{(1) SUPERB Paper}\footnote{Following the SUPERB official website \url{https://superbbenchmark.org}, we refer to the original benchmark, including 10 speech processing tasks as \textbf{SUPERB Paper} to distinguish it from other versions of the SUPERB benchmark. When discussing SUPERB, we are referring to the overall benchmark, while mentioning SUPERB Paper indicates the specific original 10-task benchmark.}~\cite{yang21c_interspeech} is a speech SSL benchmark with 10 tasks, focusing on content, speaker, semantics, and paralinguistics tasks. It adheres to the principles of lightweight fine-tuning, freezing the upstream model, and only training a downstream model defined by the task to faithfully demonstrate the generalization ability of the SSL model itself. Despite the original intention to perform lightweight fine-tuning, evaluating an SSL model even once still requires several days.

\textbf{(2) SUPERB-SG}~\cite{DBLP:conf/acl/TsaiCHHLYDLLSCH22} is an extended benchmark of SUPERB, aimed at evaluating the ability of speech SSL models in semantic and generative tasks in order to evaluate the model's abilities in various aspects more comprehensively. In contrast, we focus on the Speech Enhancement (SE) and Speech Separation (SS) tasks in the benchmark.

\textbf{(3) SUPERB Challenge}~ \cite{DBLP:conf/slt/FengDYYLSCHWCWMLL22} is comprised of 7 tasks from SUPERB paper and 3 tasks from SUPERB-SG, aiming to comprehensively evaluate the performance, generalization ability, and efficiency of speech SSL models. Our analysis shows that our results can effectively approximate the model's ranking in SUPERB Challenge.

In this work, we utilize MiniSUPERB, producing a 0.982 Spearman's rank correlation coefficient with the results presented in the SUPERB Challenge and 0.954 in the SUPERB paper, while reducing computation by over 97\%.

\section{MiniSUPERB}
We propose MiniSUPERB, a lightweight version of SUPERB, to enable researchers to quickly evaluate the performance of SSL models during the model development process as a test before full evaluation. This can help to avoid excessive computation and time costs during the model development process. To achieve this goal, we have improved MiniSUPERB's design in several aspects, including task selection, dataset reduction, downstream model simplification, and offline feature extraction. Here we provide details of MiniSUPERB's design.

\begin{table*}[h]
\centering
\caption{Details of investigated SSL representations. LibriSpeech and Libri-light are denoted as LS and LL, respectively. Mix 94k HR refers to a mix of Libri-light\cite{DBLP:conf/icassp/KahnRZKXMKLCFLS20}, VoxPopuli\cite{DBLP:conf/acl/WangRLWTHWPD20}, and GigaSpeech\cite{DBLP:conf/interspeech/ChenCWDZWSPTZJK21}. For the pretraining methods, we abbreviate ”vector quantization” as VQ, ”future” as F, ”masked” as M, ”generation” as G, ”contrastive
discrimination” as C, ”token prediction/classification” as P, ”gated relative position bias” as GREP, and ”Utterance Mixing” as UM. Parameters for both pretraining and inference are counted.}
\label{table:models}
\resizebox{\textwidth}{!}{%
\begin{tabular}{l||c|c|c|c|c|c|c }
\hline
Method            & Network         & \#Params     & Stride & Input    & Corpus     & Pretraining                        & Official Github                  \\ \hline\hline
FBANK             & -               & 0         & 10ms   & waveform & -                                           & -  & -                                \\ 
TERA              & 3-Trans         & 21.33M    & 10ms   & FBANK    & LS 960 hr  & time/freq M-G                      & s3prl / s3prl                    \\ 
DeCOAR 2.0        & 12-Trans        & 89.84M     & 10ms   & FBANK    & LS 960 hr  & time M-G + VQ                      & awslabs / speech-representations \\ 
modified CPC      & 5-Conv, 1-LSTM  & 1.84M     & 10ms   & waveform & LL 60k hr  & F-C                                & facebookresearch / CPC\_audio    \\ 
wav2vec 2.0 Base  & 7-Conv 12-Trans & 95.04M     & 20ms   & waveform & LS 960 hr  & M-C + VQ                           & pytorch / fairseq                \\ 
wav2vec 2.0 Large & 7-Conv 24-Trans & 317.38M    & 20ms   & waveform & LL 60k hr  & M-C + VQ                           & pytorch / fairseq                \\ 
HuBERT Base       & 7-Conv 12-Trans & 94.68M     & 20ms   & waveform & LS 960 hr  & M-P + VQ                           & pytorch / fairseq                \\ 
HuBERT Large      & 7-Conv 24-Trans & 316.61M    & 20ms   & waveform & LL 60k hr  & M-P + VQ                           & pytorch / fairseq                \\ 
WavLM Base        & 7-Conv 12-Trans & 94.70M     & 20ms   & waveform & LS 960 hr  & M-P + VQ + GREP + UM  & microsoft / unillm               \\ 
WavLM Base+       & 7-Conv 12-Trans & 94.70M     & 20ms   & waveform & Mix 94k hr & M-P + VQ + GREP + UM  & microsoft / unillm               \\ 
WavLM Large       & 7-Conv 24-Trans & 316.62M    & 20ms   & waveform & Mix 94k hr & M-P + VQ + GREP + UM & microsoft / unillm               \\ \hline
\end{tabular}%
}
\end{table*}
\subsection{Tasks and datasets}
We choose 4 representative tasks from SUPERB Challenge, including ASR, SID, SE, and SS.
To further reduce the cost of running MiniSUPERB, we only use 10\% of the original training dataset for each task and scale the training steps accordingly.

\textbf{Automatic Speech Recognition (ASR)} transcribes spoken words into text. It is used to assess the ability of a model to extract content information. We use Libri-light \cite{DBLP:conf/icassp/KahnRZKXMKLCFLS20} 10 hours limited-resource training set for training,  and LibriSpeech dev-clean/test-clean subsets are used for validation and testing. Word error rate (WER) is used as the evaluation metric.  We use a 256-unit 3-layer BLSTM and CTC loss as the objective.

\textbf{Speaker Identification (SID)} is a multi-class classification task that identifies the speaker of each utterance. It is used to evaluate speaker capability. The same predefined set of speakers is used for both training and testing. VoxCeleb1 \cite{DBLP:conf/interspeech/NagraniCZ17} dataset is adopted, and a 10\% subset is used for training to speed up the process. We sample 11 speech samples per speaker to create the training set. Accuracy (ACC) is used as the evaluation metric. Mean-pooling followed by a linear layer is used to predict speaker and cross-entropy loss as the objective.

% \textbf{Speaker Identification (SID)} is a multi-class classification task that identifies the speaker of each utterance. It is used to evaluate speaker capability. The same predefined set of speakers is used for both training and testing. VoxCeleb1 \cite{DBLP:conf/interspeech/NagraniCZ17} dataset is adopted, and a 10\% subset is used for training to speed up the process. We randomly sample 11 speech samples per speaker to create the training set. Accuracy (ACC) is used as the evaluation metric. Mean-pooling followed by a linear layer is used to predict speaker and cross-entropy loss as the objective.

\textbf{Speech Enhancement (SE)} aims to remove background noise from a noisy speech to improve its perceived quality and intelligibility. We use it to evaluate generative capability under noisy conditions. The Voicebank DEMAND \cite{DBLP:conf/ococosda/VeauxYK13} dataset is used, with a 10\% subset used for training. 
We divide the training data into five classes based on PESQ score intervals [~-0.5, 2.6, 3.1, 3.6, 4, 4.5], and extract 10\% of the samples from each class. The evaluation metrics are Perceptual Evaluation of Speech Quality (PESQ) and Short-Time Objective Intelligibility (STOI). 
We use a 256-unit 3-layer BLSTM to predict the spectral mask for the clean signal. The mean square error between the predicted mask and Ideal Non-negative Phase Sensitive Mask (INPSM) is used as the objective.

\textbf{Speech Separation (SS)} separates target speech from background interference and is used to evaluate the generative capability of SSL models when input is a mixture of acoustic signals. The LibriMix \cite{cosentino2020librimix} dataset is used under the 16kHz, 2 speakers, and mix\_clean setting. We sample training set to 10\% and random sample validation set from 3000 utterances to 1000 utterances. To sample the training set, we first divide the training data into 5 classes based on the SNR intervals [0, 5, 10, 15, 20, 25], and then sample 10\% of the data from each class. The evaluation metric is the scale-invariant signal-to-distortion ratio improvement (SI-SDRi). A 256-unit 3-layer BLSTM is used to predict the short-time Fourier transform (STFT) masks for each speaker, and the predictions are transformed back to the time domain using inverse short-time Fourier transform (iSTFT). The mean square error between the predicted mask and INPSM is used as the objective.

% \textbf{Speech Separation (SS)} separates target speech from background interference and is used to evaluate the generative capability of SSL models when input is a mixture of acoustic signals. The LibriMix \cite{cosentino2020librimix} dataset is used under the 16kHz, 2 speakers, and mix\_clean setting. We random sample training set to 10\% and validation set from 3000 utterances to 1000 utterances. The evaluation metric is the scale-invariant signal-to-distortion ratio improvement (SI-SDRi). A 256-unit 3-layer BLSTM is used to predict the short-time Fourier transform (STFT) masks for each speaker, and the predictions are transformed back to the time domain using inverse short-time Fourier transform(iSTFT). The mean square error between the predicted mask and INPSM 
% is used as the objective.

\subsection{Offline feature extraction}
SUPERB uses a lightweight fine-tuning approach, which involves freezing the parameters of the upstream model and training a small task-specific downstream model to focus on the robustness of the upstream model's representation. While research has shown the effectiveness of this method in evaluating model performance, the upstream model still needs to extract representations throughout the entire training process, consuming a significant amount of computing resources and time.

To address this issue, we extract the speech representations of the upstream model offline and store them on the hard disk, which can be directly used during the training of downstream models. As the representation of the original dataset can occupy an unaffordable amount of space (as shown in Table \ref{table: feature size}, the representations extracted from a Base model require over ten times more storage than the raw waveform.), we limit the storage usage by using only 10\% of the training data to ensure that all researchers can afford it. For SID, because the downstream model applies mean pooling followed by a linear layer. This allows the representation to be compacted into a smaller space by averaging along the time axis in advance.
%For SID, due to the design of the downstream model, the representation can be stored in a smaller space by taking the average along the time axis direction.

Since most of the computations are performed by the upstream model, this method trades hard disk space for a faster and more economical evaluation method.

\begin{figure}[t]
\includegraphics[width=\linewidth]{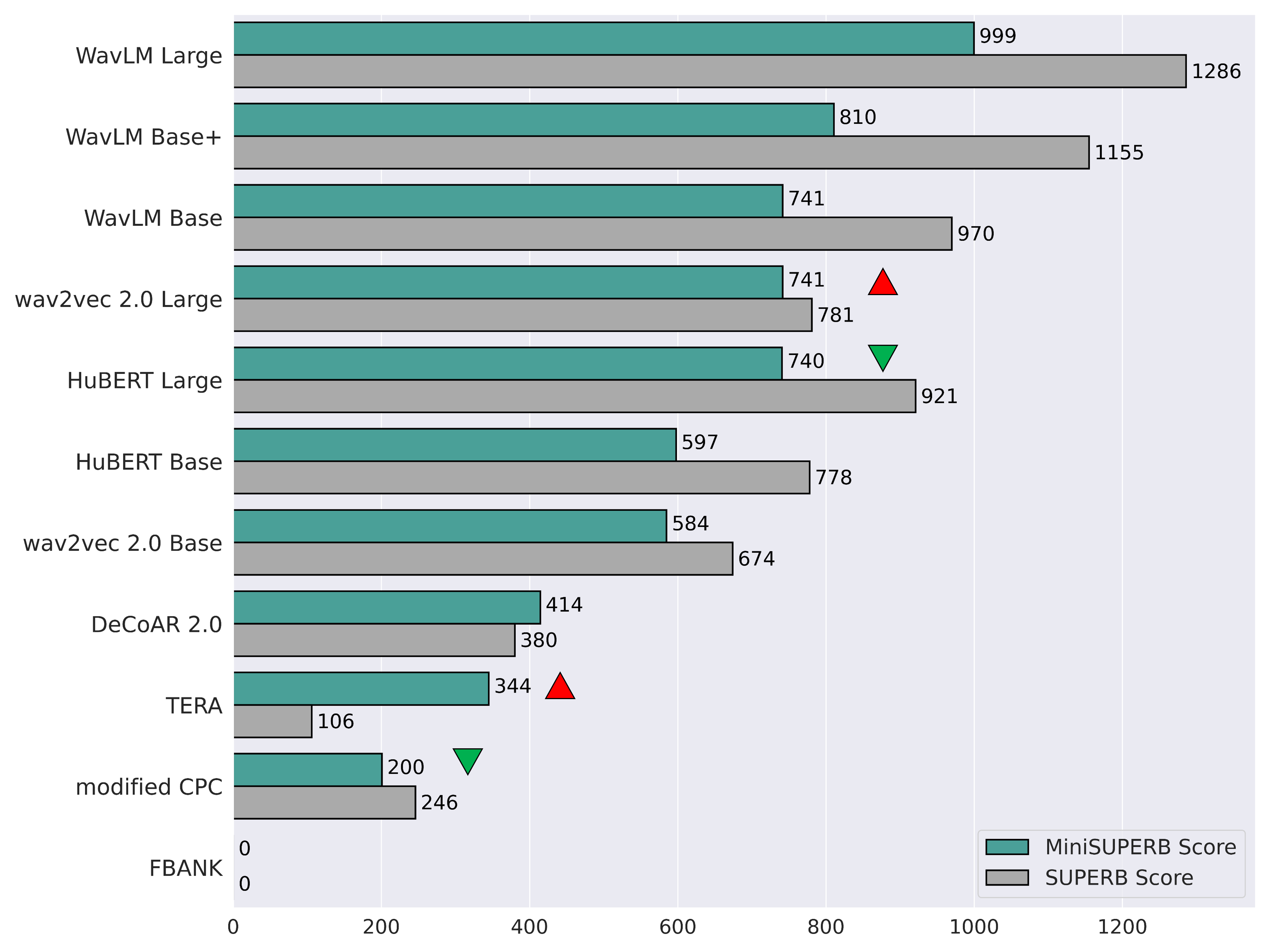}
  \caption{The Ranking of MiniSUPERB and SUPERB Challenge. This figure shows the consistency of the ranking of MiniSUPERB. Red arrows mean rank up and green arrows mean rank down in MiniSUPERB compared with SUPERB.}
  \label{fig:The Ranking of MiniSUPERB and SUPERB}
\end{figure}

\settowidth\mylenA{0.909 (SUPERB Challenge - MiniSUPERB)} 
\setlength\mylenB{(\mylenA-2\tabcolsep)/2}
\begin{table*}[t]
\caption{Evaluating various SSL representations on various downstream tasks. Results under SUPERB Challenge (left) and MiniSUPERB (right) settings are shown, with the former collected from the SUPERB official website. It also shows the change in ranking relative to the SUPERB Challenge. The numbers are collected with public-available checkpoints or codes.}
\resizebox{\textwidth}{!}{%
%\begin{tabular}{l||*{2}{wc{\mylenB}|}*{3}{wc{\mylenB}|}|r}
\begin{tabular}{l||rr|rr|rrrr|rr||r}
\hline
                 \multicolumn{1}{c||}{\multirow{2}{*}{Model}} &  \multicolumn{2}{c|}{ASR}    & \multicolumn{2}{c|}{SID}          & \multicolumn{4}{c|}{SE}     & \multicolumn{2}{c||}{SS}          & 
                 \multicolumn{1}{c}{\multirow{1}{*}{MiniSUPERB}}  \\
                 \cline{2-10}
                  &\multicolumn{2}{c|}{WER ↓}    & \multicolumn{2}{c|}{ACC ↑}     & \multicolumn{2}{c|}{PESQ ↑}     &  \multicolumn{2}{c|}{STOI ↑}      & \multicolumn{2}{c||}{Si-SDRi ↑}   &
                  \multicolumn{1}{c}{\multirow{1}{*}{Rank}}  
                     \\ \hline\hline
WavLM Large       &\textbf{3.44}& 6.94  &\textbf{95.49}& \textbf{84.74}       &\textbf{2.70}&  \multicolumn{1}{c|}{\textbf{3.02}}       &\textbf{94.49}& \textbf{95.22}        &\textbf{11.19}& \textbf{10.21}      & \textbf{1}               \\ 
WavLM Base+       &5.59& 9.64  &89.42& 61.48 &2.63& \multicolumn{1}{c|}{2.92}       &94.25& 94.82       &10.85& 9.57       & 2               \\ 
WavLM Base        &6.21& 10.39 &84.51& 58.45       &2.58& \multicolumn{1}{c|}{2.90} &94.01& 94.60 &10.37& 8.93 &     \ \ 3       \\ 
wav2vec 2.0 Large &3.75& 7.21  &86.14& 66.13 &2.52& \multicolumn{1}{c|}{2.93} &94.00& 94.80 &10.02& 7.59 & (↑ 1) \ \ 4               \\ 
HuBERT Large      &3.62& \textbf{6.87}  &90.33& 68.97       &2.64& \multicolumn{1}{c|}{2.92}       &94.20& 94.77 &10.45& 7.44 & (↓ 1) \ \ 5       \\ 

HuBERT Base       &6.42& 11.04 &81.42& 53.16       &2.58& \multicolumn{1}{c|}{2.89} &93.90& 94.68 &9.36& 6.37       & 6               \\ 
wav2vec 2.0 Base  &6.43& 10.93&75.18& 53.93      &2.55& \multicolumn{1}{c|}{2.86} &93.90& 94.44       &9.77& 6.65 & 7               \\ 
DeCoAR 2.0        &13.02& 24.65 &74.42& 42.29       &2.47& \multicolumn{1}{c|}{2.78} &93.20& 94.06 &8.54& 6.52 & 8               \\ 
TERA              &18.17& 41.22 &57.57& 38.52       &2.54& \multicolumn{1}{c|}{2.79}       &93.60& 94.31 &10.19& 6.49       & (↑ 1) \ \  9       \\ 
modified CPC      &20.18& 42.62 &39.63& 15.40       &2.57& \multicolumn{1}{c|}{2.70} &93.70& 94.06 &10.40& 6.31  & (↓ 1) 10        \\ 
FBANK             &23.18& 59.12 &20.06& 12.77      &2.55& \multicolumn{1}{c|}{2.63} &93.64& 93.65 &9.23& 5.12 & 11              \\
\hline
\end{tabular}
\label{table:model score}
}%
\end{table*}

\begin{table}[]
\caption{Spearman's $\rho$ between rankings of MiniSUPERB and SUPERB (SUPERB Paper and SUPERB Challenge). MiniSUPERB offers researchers a choice of two settings: one utilizing ASR and SID, and the other incorporating ASR, SID, SE, and SS.}
\centering
\setlength{\tabcolsep}{4pt}
\begin{tabular}{l||c|c}
\hline
                 & \multirow{2}{*}{ASR + SID} & ASR + SID\\ 
                 &            &  + SE + SS \\ \hline
SUPERB Paper \cite{yang21c_interspeech}     &   0.954      &  |  \\ \hline
SUPERB Challenge \cite{DBLP:conf/slt/FengDYYLSCHWCWMLL22} &    0.909     &    \textbf{0.982}   \\ \hline
\end{tabular}
\label{table:spearman}
\end{table}

\subsection{MiniSUPERB score}
To demonstrate the effectiveness of MiniSUPERB in approximating the SUPERB Paper and SUPERB Challenge, we ranked the upstream models using SUPERB score from SUPERB Challenge \cite{DBLP:conf/slt/FengDYYLSCHWCWMLL22} but with newly defined reference points obtained from our experiments. Here we refer to it as MiniSUPERB score for distinction.

To calculate MiniSUPERB score, all metrics should be in a higher-better manner, and each task should only have one metric.
We convert the metric of ASR from WER to WACC and calculate the arithmetic mean of PESQ and STOI from SE. We describe MiniSUPERB score of upstream model $u$ as follows: 
\begin{align}
    minisuperb_s(u) = \frac{1000}{|T|}\frac{s_t(u)-s_t(baseline)}{s_t(SOTA)-s_t(baseline)}
\end{align}
where $s_t$ is the metric for task $t$, $s_t(u)$ is the corresponding metric of upstream model $u$, and $T$ is the set of all four tasks. Here we use FBANK as the baseline and SOTA to be the best SSL model we evaluate in each task.

\begin{table*}[t]
\caption{MACs comparison between SUPERB Challenge and MiniSUPERB, and the reduction ratio of each task}
\centering
\resizebox{\textwidth}{!}{%
\begin{tabular}{l|ccccc|ccccc|c }
\hline
\multicolumn{1}{c|}{\multirow{2}{*}{\textbf{Model}}} & \multicolumn{5}{c|}{SUPERB Challenge}                                                                                                                                 & \multicolumn{5}{c|}{MiniSUPERB}                                                                                                                                             & \multirow{2}{*}{\textbf{Reduction (\%)}} \\  \cline{2-11}                     
                       & \multicolumn{1}{c|}{ASR}        & \multicolumn{1}{c|}{SID}        & \multicolumn{1}{c|}{SE}         & \multicolumn{1}{c|}{SS}         & \textbf{Total}      & \multicolumn{1}{c|}{ASR}        & \multicolumn{1}{c|}{SID}        & \multicolumn{1}{c|}{SE}         & \multicolumn{1}{c|}{SS}         & \multicolumn{1}{c|}{\textbf{Total}} & \\ \hline
                       
WavLM Large            & \multicolumn{1}{c|}{8.7E+17}  & \multicolumn{1}{c|}{1.3E+18}  & \multicolumn{1}{c|}{4.3E+17}  & \multicolumn{1}{c|}{6.5E+17}  & \textbf{3.3E+18}  & \multicolumn{1}{c|}{1.2E+16}  & \multicolumn{1}{c|}{6E+16}  & \multicolumn{1}{c|}{4.7E+15}  & \multicolumn{1}{c|}{6.1E+15}  & \textbf{8.2E+16}      & \textbf{97.4680}                           \\ 
WavLM Base+            & \multicolumn{1}{c|}{3.4E+17}  & \multicolumn{1}{c|}{5.0E+17}  & \multicolumn{1}{c|}{1.7E+17}  & \multicolumn{1}{c|}{2.5E+17}  & \textbf{1.3E+18}  & \multicolumn{1}{c|}{4.7E+15}  & \multicolumn{1}{c|}{2.3E+16}  & \multicolumn{1}{c|}{1.8E+15}  & \multicolumn{1}{c|}{2.4E+15}&
\textbf{3.2E+16}                  & \textbf{97.4685}                           \\ 
WavLM Base             & \multicolumn{1}{c|}{3.4E+17}  & \multicolumn{1}{c|}{5.0E+17}  & \multicolumn{1}{c|}{1.7E+17}  & \multicolumn{1}{c|}{2.5E+17}  & \textbf{1.3E+18}  & \multicolumn{1}{c|}{4.7E+15}  & \multicolumn{1}{c|}{2.3E+16}  & \multicolumn{1}{c|}{1.8E+15}  & \multicolumn{1}{c|}{2.4E+15}  & \textbf{3.2E+16}                  & \textbf{97.4685}                           \\ 
wav2vec 2.0 Large      & \multicolumn{1}{c|}{8.7E+17}  & \multicolumn{1}{c|}{1.3E+18}  & \multicolumn{1}{c|}{4.3E+17}  & \multicolumn{1}{c|}{6.5E+17}  & \textbf{3.3E+18}  & \multicolumn{1}{c|}{1.2E+16}  & \multicolumn{1}{c|}{6E+16}  & \multicolumn{1}{c|}{4.7E+15}  & \multicolumn{1}{c|}{6.1E+15}  & \textbf{8.2E+16}                  & \textbf{97.4680}                           \\ 
HuBERT Large           & \multicolumn{1}{c|}{8.7E+17}  & \multicolumn{1}{c|}{1.3E+18}  & \multicolumn{1}{c|}{4.3E+17}  & \multicolumn{1}{c|}{6.5E+17}  & \textbf{3.3E+18}  & \multicolumn{1}{c|}{1.2E+16}  & \multicolumn{1}{c|}{6E+16}  & \multicolumn{1}{c|}{4.7E+15}  & \multicolumn{1}{c|}{6.1E+15}  & \textbf{8.2E+16}                  & \textbf{97.4680}                           \\ 
HuBERT Base            & \multicolumn{1}{c|}{3.4E+17}  & \multicolumn{1}{c|}{5.0E+17}  & \multicolumn{1}{c|}{1.7E+17}  & \multicolumn{1}{c|}{2.5E+17}  & \textbf{1.3E+18}  & \multicolumn{1}{c|}{4.7E+15}  & \multicolumn{1}{c|}{2.3E+16}  & \multicolumn{1}{c|}{1.8E+15}  & \multicolumn{1}{c|}{2.4E+15}  & \textbf{3.2E+16}                  & \textbf{97.4685}                           \\ 
wav2vec 2.0 Base       & \multicolumn{1}{c|}{3.4E+17}  & \multicolumn{1}{c|}{5.0E+17}  & \multicolumn{1}{c|}{1.7E+17}  & \multicolumn{1}{c|}{2.5E+17}  & \textbf{1.3E+18}  & \multicolumn{1}{c|}{4.7E+15}  & \multicolumn{1}{c|}{2.3E+16}  & \multicolumn{1}{c|}{1.8E+15}  & \multicolumn{1}{c|}{2.4E+15}  & \textbf{3.2E+16}                  & \textbf{97.4685}                           \\ 
modified CPC           & \multicolumn{1}{c|}{5.0E+16} & \multicolumn{1}{c|}{6.1E+16} & \multicolumn{1}{c|}{2.1E+16} & \multicolumn{1}{c|}{3.1E+16} & \textbf{1.6E+17} & \multicolumn{1}{c|}{6.5E+14} & \multicolumn{1}{c|}{2.8E+15} & \multicolumn{1}{c|}{2.7E+14} & \multicolumn{1}{c|}{3.7E+14} & \textbf{4.1E+15}                 & \textbf{97.4847}                          \\ 

DeCOAR 2.0             & \multicolumn{1}{c|}{2.3E+17} & \multicolumn{1}{c|}{3.3E+17} & \multicolumn{1}{c|}{1.1E+17} & \multicolumn{1}{c|}{1.7E+17} & \textbf{8.5E+17} & \multicolumn{1}{c|}{3.2E+15} & \multicolumn{1}{c|}{1.5E+16} & \multicolumn{1}{c|}{1.3E+15} & \multicolumn{1}{c|}{1.6E+15} & \textbf{2.1E+16}                 & \textbf{97.4699}                          \\ 
TERA                   & \multicolumn{1}{c|}{1.2E+17} & \multicolumn{1}{c|}{1.7E+17} & \multicolumn{1}{c|}{5.7E+16} & \multicolumn{1}{c|}{8.6E+16} & \textbf{4.4E+17} & \multicolumn{1}{c|}{1.7E+15} & \multicolumn{1}{c|}{7.8E+15} & \multicolumn{1}{c|}{6.7E+14} & \multicolumn{1}{c|}{8.9E+14} & \textbf{1.1E+16}                 & \textbf{97.4718}                          \\ 

FBANK                  & \multicolumn{1}{c|}{9.0E+15} & \multicolumn{1}{c|}{1.4E+14} & \multicolumn{1}{c|}{4.8E+14} & \multicolumn{1}{c|}{8.5E+14} & \textbf{1.0E+16} & \multicolumn{1}{c|}{9.2E+13} & \multicolumn{1}{c|}{6.6E+12} & \multicolumn{1}{c|}{4.4E+13} & \multicolumn{1}{c|}{7.8E+13} & \textbf{2.2E+14}                 & \textbf{97.8952}                          \\ \hline
\end{tabular}%
\label{table: compute_cost}
}
\end{table*}

\begin{table}[t]
\caption{The space usage of waveform and upstream model representation, where usage under SUPERB (left) and MiniSUPERB (right) settings are shown. The size is measured in GigaBytes (GB).}
\centering
\setlength{\tabcolsep}{2pt}
\begin{tabular}{c|c|c|c|c}
\hline
             & ASR        & SID        & SE          & SS          \\ \hline\hline
waveform     & 7 / 1.3    & 38.7 / 7 & 2.2 / 0.464   & 5.7 / 1.1   \\ \hline
FBANK        & 35.6 / 6.8 & 0.6 / 0.115  & 7.1 / 1.6   & 37.4 / 7.6  \\ \hline
TERA         & 456 / 87   & 2.4 / 0.454  & 42.6 / 9.5  & 227 / 46    \\ \hline
DeCoAR 2.0   & 734 / 140  & 6.6 / 1.3 & 66.6 / 15 & 361 / 72    \\ \hline
modified CPC & 61 / 14.4  & 0.6 / 0.115  & 10.6 / 2.4  & 55.4 / 12 \\ \hline
Base Model   & 734 / 140  & 6.6 / 1.3  & 66.6 / 15 & 361 / 72    \\ \hline
Large Model  & 1866 / 354 & 16.6 / 3 & 173 / 37    & 871 / 177   \\ \hline
\end{tabular}
\label{table: feature size}
\end{table}

\subsection{Computational cost evaluation}
To evaluate the efficiency of the proposed improvement over the original SUPERB, we measured the computational cost of both SUPERB and MiniSUPERB on four different downstream tasks. Specifically, we report the estimated number of multiply-accumulate operations (MACs) required to train the downstream models on these tasks. For each task, we describe the computational cost (in MACs) of the original SUPERB benchmark and MiniSUPERB as follows:
\begin{align}
    C_{superb} &= C_U \times S_t^{s} + C_D \times (1 + R) \times S_t^{s} \\
    C_{minisuperb} &= C_U \times S_f + C_D \times (1 + R) \times S_t^{m},
\end{align}
where $C_U$ is the forward MACs of the given upstream model, $C_D$ is the forward MACs of the given downstream model, $S_t^{s}$ and $S_t^{m}$ are the number of training steps of SUPERB and MiniSUPERB respectively, $S_f$ denotes the number of steps required for an upstream model to extract features over the dataset, and $R = \frac{\textrm{Operations per backward pass}}{\textrm{Operations per forward pass}}$ is the backward-forward computation ratio. We set $R=2$, for it has shown a reasonable ratio to estimate the training cost in the literature~\cite{DBLP:conf/ijcnn/SevillaHHBHV22}. Notably, in MiniSUPERB, we perform offline feature extraction, and the upstream model only needs to forward once regarding one data point. Typically, $S_f$ is much smaller than $S_t$, resulting in significant computation improvements.

\subsection{Self-supervised models}
We evaluate MiniSUPERB on 11 well-known upstream models, including TERA \cite{DBLP:journals/taslp/LiuLL21}, DeCoAR 2.0 \cite{DBLP:journals/corr/abs-2012-06659}, Modified CPC \cite{DBLP:conf/icassp/RiviereJMD20}, wav2vec 2.0 \cite{DBLP:conf/nips/BaevskiZMA20}, HuBERT \cite{DBLP:journals/taslp/HsuBTLSM21}, and WavLM \cite{DBLP:journals/jstsp/ChenWCWLCLKYXWZ22}. Following SUPERB, FBANK is also the baseline. For the baseline in SID, FBANK does not employ CMVN (Cepstral Mean and Variance Normalization), as the input mean and variance are important information for SID, and using CMVN can significantly hurt its performance. The detailed properties of upstream models are shown in Table \ref{table:models}.

\begin{table}[t]
\caption{Results on different tasks with different amounts of training data}
\centering
\label{table: ablation}
\resizebox{\textwidth/2-10pt}{!}{%
\begin{tabular}{lrrrrr}
\hline
\multirow{2}{*}{\textbf{Model}} & \multicolumn{1}{c}{ASR} && \multicolumn{1}{c}{SID} && \multicolumn{1}{c}{SS}\\ \cline{2-2}\cline{4-4}\cline{6-6} 
 &WER ↓  && ACC ↑  && Si-SDRi ↑ \\ \hline
\multicolumn{5}{l}{\textit{10\%}} \\\hline
FBANK        &59.12 && 12.77  &&5.12  \\ 
TERA         &41.22 &&38.52  &&6.49     \\ 
modified CPC &42.62  &&15.40 &&6.31  \\ 
wav2vec 2.0 Base&10.93 &&53.93  &&6.65    \\ 
HuBERT Base&11.04 &&53.16 &&6.37    \\ 
WavLM  Base&10.39 && 58.45  &&8.93    \\ 
WavLM  Base+&\textbf{9.64} &&\textbf{61.48} &&\textbf{9.57}    \\ 
\hline
\multicolumn{5}{l}{\textit{1\%}} \\\hline
FBANK        &85.50&&2.34   &&0.003  \\ 
TERA         &75.07&&2.04   &&2.86     \\ 
modified CPC &66.89&&3.18&&1.76  \\ 
wav2vec 2.0 Base&22.97&&\textbf{4.13}&&2.04    \\ 
HuBERT Base&22.49&&3.02&&1.77    \\ 
WavLM  Base&21.98&&2.51&&5.44    \\ 
WavLM  Base+&\textbf{20.32}&&3.62 &&\textbf{5.82}    \\ 
\hline
\end{tabular}
}

\end{table}

\section{Experimental setup}
Following SUPERB, we freeze the parameters of the upstream model in all downstream tasks. However, we further separate the computation of the upstream model and the training of the downstream model to increase training efficiency. The upstream model only extracts representations for all speech data offline and does not participate in the training of the downstream model. During the training of the downstream model, we directly load representations for each hidden layer of the upstream model. We then perform layer normalization and weighted sum on the representations, which serve as the input to the downstream model. Here the weights are trainable parameters that are jointly trained with the downstream model.

To evaluate computational costs, we follow the setting in SUPERB Challenge and rely on Microsoft DeepSpeed\footnote{\url{https://github.com/microsoft/DeepSpeed}}  to estimate the forward MACs of both the upstream and downstream models. Following SUPERB Challenge, we perform this evaluation over the same 32 utterances sampled from the LibriSpeech test-clean dataset as inputs. When performing the offline feature extraction, we set the batch size to 1 to bypass padding and normalization issues for simplicity.

\section{Result}

We present the performance and ranking of each model in Table \ref{table:model score} and Figure \ref{fig:The Ranking of MiniSUPERB and SUPERB}. It shows that WavLM Large almost outperforms all other models in all four tasks in both SUPERB and MiniSUPERB, followed by HuBERT and wav2vec 2.0. Overall, the rankings of MiniSUPERB and SUPERB Challenge are similar, with only a few differences observed. Specifically, the rankings for HuBERT Large and WavLM Base, as well as modified CPC and TERA, are swapped between the two models. Despite observing changes in rankings, their scores are still quite close. We believe that this is mainly due to dataset sampling and simplification of downstream models, as we trade a small degree of ranking precision for reduced computational cost.

Table \ref{table:spearman} shows that the Spearman's $\rho$ between ASR and SID scores and the SUPERB Paper Leaderboard is 0.954, and the correlation with the SUPERB Challenge Leaderboard after calculating all four tasks is 0.982, indicating that MiniSUPERB partially approximates the SUPERB Challenge leaderboard.

As shown in Table \ref{table: compute_cost}, MiniSUPERB has reduced MACs by about 97\% across the four selected tasks. Since we first perform feature extraction on the speech data before training downstream models, the upstream model only needs to infer the dataset once, and both the dataset and the number of training iterations are reduced by about 90\%. The downstream models for ASR, SS, and SE are also simplified to a three-layer LSTM to reduce complexity, resulting in a significant reduction in the overall MACs required to run MiniSUPERB.

Before training the downstream tasks, we extract the features and store them on disk, with different sizes depending on the dataset and model. As models of the same size extract representations of the same size, we unify wav2vec 2.0 Base, HuBERT Base, WavLM Base, and WavLM Base+ as \textbf{Base model}, and wav2vec 2.0 Large, HuBERT Large, and WavLM Large as \textbf{Large model}. We show the space occupied by the representations of each MiniSUPERB model in Table \ref{table: feature size}, along with the space occupied by the original speech data for reference. Large models require the most space for ASR, with 354 GB, while SID requires the least, with only 3 GB, which is within an acceptable range.

To further examine the model's generalization ability, we sampled 1\% of the data to train the model. The results are shown in Table \ref{table: ablation}. It shows a change in the ranking of ASR due to a performance drop in TERA. However, the rankings of other models have remained unchanged. On the other hand, for SS, the model rankings are no longer consistent with the original SUPERB rankings. At the same time, WavLM Base+ demonstrates stronger generalization capabilities compared to other models. With only a small amount of training data, it remains competitive across all three tasks, whereas most other models fail on SS.

\section{Conclusions}
We present MiniSUPERB, a lightweight benchmark with automatic speech recognition, speaker identification, speech separation, and speech enhancement, aimed at quickly evaluating the performance of speech SSL models on the original SUPERB and SUPERB-SG.
We evaluated 11 SSL models and demonstrated through detailed analysis that the model rankings in MiniSUPERB are highly positively correlated with the original SUPERB benchmark (SUPERB Paper) and SUPERB Challenge with Spearman's correlation of 0.954 and 0.982 respectively. We also reduced computational costs by around 97\% of MACs on ASR, SID, SE, and SS.
To the best of our knowledge, MiniSUPERB is the first work that considers the computational cost of evaluating a model on downstream tasks. MiniSUPERB is also the first work that serves as a proxy benchmark for multiple benchmarks.
We will open-source all the code and encourage researchers to use MiniSUPERB for a quick evaluation. By doing so, researchers can avoid unnecessary computational and time costs before using SUPERB for a complete evaluation.

\section{Acknowledgement}
We thank the National Center for High-performance Computing (NCHC) of National Applied Research Laboratories (NARLabs) in Taiwan for providing computational and storage resources.

\bibliographystyle{IEEEbib}
\bibliography{refs}

\end{document}